\documentstyle[preprint,prl,aps]{revtex}
\begin{document}
\draft
\title{Quantum noise limits for nonlinear, phase-invariant amplifiers}{}{}
\author{Dmitri Kouznetsov\thanks{E-mail: kusnecov@aleph.cinstrum.unam.mx}}
\address{Centro de Instrumentos, Universidad Nacional
Aut\'onoma de M\'exico, Ap.70-186, \hbox{Cd. Universitaria}, 04510
D. F., Mexico,~ and~
Lebedev Physics Institute, Leninsky 54, 117924 Russia}
\author{Roberto Ortega}
\address{Centro de Instrumentos, Universidad Nacional
Aut\'onoma de M\'exico, Ap.70-186, \hbox{Cd. Universitaria}, 04510
D. F., Mexico}
\author{Daniel Rohrlich\thanks{E-mail: daniel@vm.tau.ac.il}}
\address{School of Physics and Astronomy,
Tel-Aviv University, Ramat-Aviv 69978 Tel-Aviv, Israel}
\date{\today}
\maketitle
\begin{abstract}
Any quantum device that amplifies coherent states of a field while
preserving their phase generates noise.  A nonlinear, phase-invariant
amplifier may generate less noise, over a range of input field strengths,
than any linear amplifier with the same amplification.  We present
explicit examples of such nonlinear amplifiers, and derive lower bounds
on the noise generated by a nonlinear, phase-invariant quantum amplifier.
\end{abstract}
Any quantum device that amplifies a field while preserving its phase
generates noise.  Noise is unavoidable since the phase and number
operators for modes of a quantum field do not commute.  Two values
characterize a phase-preserving or phase-invariant\cite{note} amplifier:
the amplification coefficient
$G = \langle A \rangle / \langle a \rangle$,
and the noise
$D = \langle  A^\dagger  A \rangle -
\langle  A^\dagger  \rangle \langle A \rangle .$
Here we focus on a single field mode, for simplicity:  $a$ and $A$
denote the field mode before and after amplification, respectively.
We take expectation values
in a coherent state $\vert \alpha \rangle$; since
$a \vert \alpha \rangle = \alpha \vert
\alpha \rangle$, the noise in the unamplified field is zero.
For a {\it linear} quantum amplifier, $G$ is a constant,
independent of the initial state, and the minimal noise is well
known\cite{haus,caves,yama}:
$D \ge |G|^2-1$ for $|G| \geq 1 .$
For a nonlinear amplifier, $G$ depends on $x\equiv \alpha^* \alpha$
(but not on the phase of $\alpha$).
In this Letter, we derive lower limits for the quantum noise in
nonlinear quantum amplifiers, and demonstrate that
a nonlinear amplifier may generate less noise, for the same
amplification, than any linear amplifier, over a range of input field
strengths.

     What makes an amplifier a {\it quantum} amplifier is that the mode
operators before and after amplification must be related by a unitary
transformation, $A=U^\dagger a U$.  Thus, $[a, a^\dagger ] = 1$
implies\cite{haus} that
$[A, A^\dagger ]=1$.  In principle, $U$ could depend on
the field operators $a$ and $a^\dagger$ alone.  However, then
there will be no amplification.  If $U$ depends on $a$ and $a^\dagger$
alone, then phase invariance requires it to depend only on $a^\dagger
a$; the noise, $\langle U^\dagger a^\dagger a U \rangle - \langle
U^\dagger a^\dagger U\rangle \langle U^\dagger a U\rangle = \vert \alpha
\vert^2 - \vert \alpha G(x)\vert^2$, cannot be negative, so $|G|\le 1$.
When $|G(x)|=1$ there is no noise, but the amplifier does not amplify.
We must introduce operators for the amplifier, and these induce
noise.  The choice $U=\exp(-iHt)~$, with $H= i(a^\dagger  b^\dagger -ab)~$,
leads to a linear amplifier.  (The amplifier degree of freedom $b$ obeys
$[b ,b^\dagger ]=1$, and $t$ is real.)   We find $A=U^\dagger aU =
a \cosh t + b^\dagger \sinh t$; for an amplifier prepared in
the ground state $G=\cosh t$, and the noise $D=\sinh^2
t$ saturates the bound $D\le \vert G \vert^2 -1$ for linear amplifiers.

     For nonlinear amplifiers, we have no general expression
for $U$ and $A$ in terms of field and amplifier degrees of freedom.
Nevertheless, we have the following lower bounds for the noise of a
nonlinear, phase-invariant quantum amplifier:

{\em Theorem.}~ Let~
$D = \langle \alpha \vert A^\dagger A\vert \alpha \rangle -
\langle \alpha \vert U^\dagger A^\dagger U\vert \alpha \rangle
\langle \alpha \vert U^\dagger A U\vert \alpha \rangle$, and
\begin{equation}
E(x)=\sum_{n=1}^\infty \frac{x^{n+1}}{n!}
\left| \frac{{\rm d}^n}{{\rm d}x^n} G(x) \right|^2~,
{}~F(x)= \sum_{n=1}^\infty  {{x^{n-1}}\over {n!}}
\left| \frac{{\rm d}^n}{{\rm d}x^n} \left( xG(x) \right) \right|^2-1~;~~
\end{equation}
Then $D \geq E(x)$ and $D \geq F(x)$~.

     {\em Proof.}  Define the set of states $\vert \alpha^{(n)} \rangle$
\begin{equation}
\vert \alpha^{(n)} \rangle
= {{(a^\dagger -\alpha^* )^n} \over {\sqrt{n!}}} \vert
\alpha \rangle~~~,
\label{ortho}
\end{equation}
where $\langle \alpha^{(m)} \vert \alpha^{(n)} \rangle
=\delta_{mn}$ and $\vert \alpha^{(0)} \rangle =\vert \alpha \rangle$.
Differentiating the expansion
\begin{equation}
  |\alpha\rangle =
e^{- \alpha^* \alpha / 2} \sum_{n=0}^{\infty}
{  {(\alpha a^\dagger)^n} \over {n!} } |0 \rangle~~~,
\label{alpha}
\end{equation}
we obtain
\begin{equation}
{\partial \over {\partial\alpha}} |\alpha\rangle =
 - { {\alpha^*} \over {2}} |\alpha\rangle + a^\dagger |\alpha\rangle ~~,
{}~{{\partial} \over {\partial\alpha^*}}  |\alpha\rangle =
- {{\alpha} \over {2}}  |\alpha\rangle ~~.
\end{equation}
Note that $A=U^\dagger aU$ does
not depend on $\alpha$, since the amplifier cannot anticipate the input state.
Thus, Eqs.\ (\ref{ortho}-\ref{alpha}) imply
\begin{equation}
{1 \over {\sqrt{n!}}} {{\partial^n}\over {\partial \alpha^n}} (\alpha G )
= \langle \alpha \vert A \vert \alpha^{(n)} \rangle~~,
{}~{1 \over {\sqrt{n!}}} {{\partial^n}\over {\partial \alpha^n}} (\alpha^* G^*
)
= \langle \alpha \vert A^\dagger \vert \alpha^{(n)} \rangle.
\end{equation}
Consider the identity operator $I_0$ in the Hilbert space of states of the
field:
\begin{equation}
I_{0}
=\sum_{n=0}^\infty \vert \alpha^{(n)} \rangle \langle \alpha^{(n)} \vert
{}~~.
\end{equation}
$I_0$ is only part of the identity operator $I$ in the Hilbert
space of states of the field and amplifier:
\begin{equation}
I =\sum_{m,n=0}^\infty \vert m, \alpha^{(n)} \rangle \langle m,
\alpha^{(n)} \vert
{}~~.
\end{equation}
The index $m$ refers to the states of the amplifier.  Without loss of
generality, we may identify $I_0$ with the $m=0$ term in $I$, where
$m=0$ represents the initial state of the amplifier.  Inserting
$I$ into the expectation value $\langle \alpha \vert A^\dagger A \vert
\alpha \rangle$ --- where we identify $\vert \alpha \rangle$ with
$\vert 0, \alpha \rangle$ --- we obtain
\begin{equation}
\langle \alpha \vert A^\dagger A \vert \alpha \rangle
=\langle \alpha \vert A^\dagger I A \vert \alpha \rangle
\ge \langle \alpha \vert A^\dagger I_0 A \vert \alpha \rangle
{}~~,
\end{equation}
\begin{equation}
D \geq  \langle \alpha \vert A^\dagger I_0 A \vert \alpha \rangle -
\vert \alpha G \vert^2 =
\sum_{n=1}^\infty \frac{1}{n!}
\left|{{\partial^n}\over {\partial \alpha^n}} (\alpha^* G^* )
\right|^2 =E(x)~~.
\end{equation}
On the other hand, since $A^\dagger A$ = $AA^\dagger -1$
and $\langle \alpha \vert AIA^\dagger\vert \alpha \rangle
\geq \langle \alpha \vert AI_0 A^\dagger \vert \alpha \rangle$,
we have
\begin{equation}
D \geq \langle \alpha \vert A I_0 A^\dagger \vert \alpha \rangle -1 -
\vert \alpha G \vert^2
=\sum_{n=1}^\infty {1\over {n!}} \left|
\frac{{\partial}^n}{{\partial} \alpha^n}
(\alpha G ) \right|^2-1~=F(x)~.
\end{equation}
{\em (End of proof)}

     Both $E(x)$ and $F(x)$ are lower bounds on the
noise of a nonlinear, phase-invariant quantum amplifier.
When $G$ is constant, $F(x)$ reduces to the bound for linear amplifiers,
$|G|^2 -1$.  But both bounds can be less than $|G(x)|^2 -1$, for a
range of values of $x$, if $|G(x)|$ decreases with $x$.
($|G(x)|$ decreasing with $x$ describes amplifier saturation.)
We have not yet shown that, for
any given $G(x)$, a nonlinear amplifier can realize these
lower bounds.  A class of amplifiers which do realize the
lower bound $F(x)$ have a linear
amplification followed by a nonlinear refraction
depending only on $a^\dagger a$:
\begin{equation}
A = e^{itH} \left[ G_0 a + (|G_0|^2 -1)^{1/2} ~b^\dagger \right] e^{-itH}
{}~~,
\end{equation}
where $t$ is real.  For example, let
$H=a^\dagger a (a^\dagger a -1)/2$. We find
(using Eq.\ (\ref{alpha})) that
$G(x)=G_0 \exp(-qx)$, and
$D =|G_0|^2 \left[ 1+x-x\exp(-|q|^2x) \right] - 1$~,~
where $q=1-e^{-it}$.
The amplifier realizes the lower bound $F(x) =D$, while
$E(x) =D-1+|G_0|^2$~.
For this class of amplifiers, the linear lower bound $D \ge |G(x)|^2-1$
also holds. It is broken in the next example.

\begin{figure}
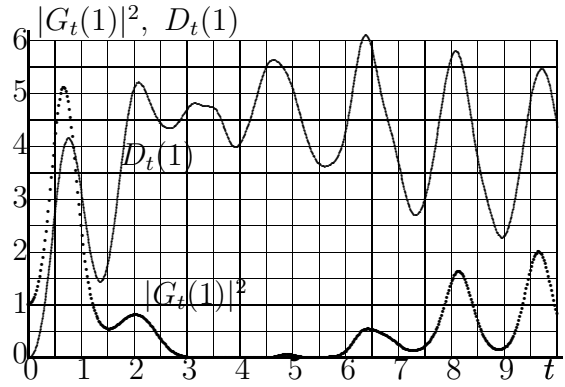

\caption{Amplification coefficient $g=G_t^2$ and noise $D_t$ versus $t$
at $x=1$ for the Cummings-Tavis amplifier with $N=10$ atoms.
At $t=0$, $G^2 =1$ (thick line) and $D=0$ (thin line).}
\label{fig1}
\end{figure}

\begin{figure}
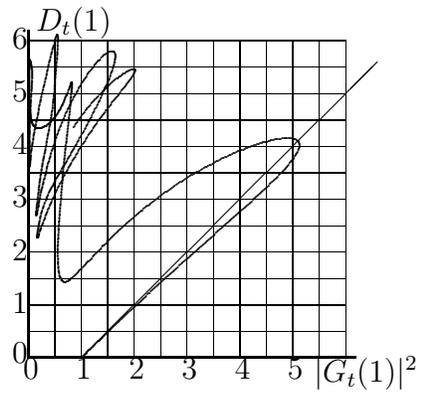

\caption{The noise $D_t$ versus $G_t^2$ under the same conditions.}
\label{fig2}
\end{figure}

\begin{figure}
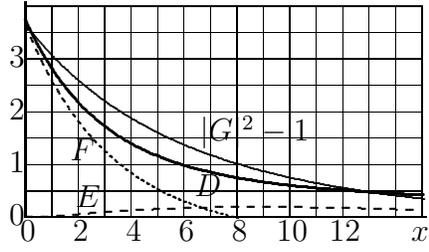

\caption{The noise $D$ (thick curve), linear lower bound $G^2-1$ (thin
curve), and lower bounds $E$ (dotted curve) and $F$ (dashed curve)
versus $x$ for the Cummings-Tavis amplifier at $N=10$, $t= 0.5$.}
\label{fig3}
\end{figure}

\begin{figure}
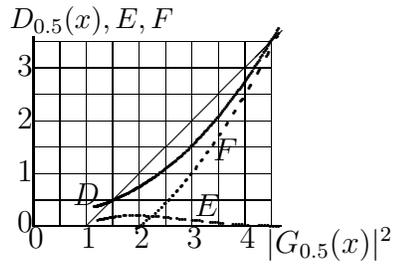

\caption{ The noise $D$ and all bounds versus $G^2$ under the same
conditions.}
\label{fig4}
\end{figure}

The resonant interaction of $N$ identical two-level atoms with a
single-mode field can be described\cite{tavis} by the Hamiltonian
$H=iab^\dagger-ia^\dagger b~~ $, where
$
b=\sum_{k=1}^N b_k~,
b_k b_j = (1-\delta_{jk}) b_j b_k ~,~~
b_k^\dagger b_j = b_j b_k^\dagger {\rm ~~ for~~} j \neq k~$, and
$b_k^\dagger b_k + b_k b_k^\dagger = 1 .
$
($b_k^\dagger$ is the operator for exciting the $k$-th atom.)
The evolution operator $\exp(-iHt)$ defines the transformation $U$.  Let
all atoms be excited in the initial state of the amplifier.
In this case $G$ is real. For $N=10$ and a given value of $t$,
let $G_t$ denote the amplification factor and $D_t$ the noise.
We compute the evolution from the initial state numerically.
Fig.\ \ref{fig1}
shows the coefficient $g=G_t^2$ and the noise $D_t$ versus $t$
at input intensity $x=1$. The noise $D_t$ oscillates
quasiperiodically while $G_t$ has ``revivals''.
Fig.\ \ref{fig2}, a plot of $D_t (1)$ versus $| G_t (1)|^2$,
shows that the linear relation $D \geq |G|^2 - 1$ is valid only at small
$t$ ~(for $G_t^2 \lesssim 2$), where the amplifier is linear, and violated
at larger $x$. Fig.\ \ref{fig3} compares $D$,
$|G|^2-1$, $E$, and $F$ as
functions of $x$, while Fig.\ \ref{fig4} is a plot of the three bounds versus
$|G(x)|^2$, for $t= 0.5$. The nonlinear amplifier beats the linear limit,
while the lower limits of the Theorem hold.

     The final example involves a small nonlinear perturbation
on a linear amplifier.  We define
\begin{equation}
A=Ca +Sb^\dagger -\epsilon \left[ a^\dagger a^2 -(C/S) a^2 b + (2C/S)
a^\dagger a b^\dagger \right]  ~~~,
\end{equation}
where $C\equiv \cosh t$, $S\equiv \sinh t$, and $[ a,a^\dagger ]=1
=[b,b^\dagger ]$, $[a,b^\dagger ]=0=[a,b]$.  Let $\epsilon$ be small and
positive.  We find $[A,A^\dagger ]=1$ to order $\epsilon$, and
$\langle A \rangle = (C-\epsilon x)\alpha$, i.e. $G=C-\epsilon x$.
The noise is $D=C^2 -1 -4Cx\epsilon <G^2 -1$, violating the linear limit
to first order in $\epsilon$.  It saturates the bound $F(x)$, while
$E(x)=0$ to this order.

In summary, we have obtained lower bounds
on the irreducible noise of a nonlinear, phase-preserving quantum
amplifier, in terms of the amplification factor $G$ and its derivatives.
For $G$ constant, the amplifier is linear,
and the bound $F$ reproduces the linear bound.
For $G$ not a constant, however, the nonlinear
bound can undercut the linear bound.  Explicit examples confirm that
the noise of a nonlinear amplifier
can be less than that of an ideal linear amplifier with the same
amplification coefficient, over a range of input field strengths.
However, we have not determined in general which functions $G(x)$
correspond to nonlinear amplifiers, nor whether the lower bounds can be
realized.

\acknowledgments
The work of D. K. was partially supported by CONACyT of Mexico.
D. R. thanks the Ticho Fund for support.

\newpage
\input{afig1}

\newpage
\input{afig2}

\newpage
\vspace{2cm}
\hspace{2cm}
\begin{picture}(160,130)
\put(-.3,-.3){\rule{.5pt}{82pt}}
\put(-.3,-.3){\rule{152pt}{.5pt}}
\multiput(-0.,9.95)(0,10){ 8}{\rule{150pt}{.1pt}}
\multiput(9.95,0. )(10,0){ 15}{\rule{.1pt}{80pt}}
\put(-10, 57){ 3}
\put(-10, 37){ 2}
\put(-10, 17){ 1}
\put(-10, -3){ 0}
\put(-2,-9){0}
\put(18,-9){2} \put(38,-9){4}
\put(58,-9){6} \put(78,-9){8}
\put(95,-9){10} \put(115,-9){12}
\put(146,-9){$x$}
\multiput(-0.8,72.5)(0.33,-0.42){3}{\tiny.}
\multiput(0.2,71.2)(0.33,-0.41){3}{\tiny.}
\multiput(1.2,69.9)(0.33,-0.41){3}{\tiny.}
\multiput(2.2,68.7)(0.33,-0.40){3}{\tiny.}
\multiput(3.2,67.5)(0.33,-0.39){3}{\tiny.}
\multiput(4.2,66.4)(0.33,-0.38){3}{\tiny.}
\multiput(5.2,65.2)(0.33,-0.37){3}{\tiny.}
\multiput(6.2,64.1)(0.33,-0.37){3}{\tiny.}
\multiput(7.2,63.0)(0.33,-0.36){3}{\tiny.}
\multiput(8.2,61.9)(0.33,-0.35){3}{\tiny.}
\multiput(9.2,60.9)(0.33,-0.35){3}{\tiny.}
\multiput(10.2,59.8)(0.33,-0.34){3}{\tiny.}
\multiput(11.2,58.8)(0.33,-0.33){3}{\tiny.}
\multiput(12.2,57.8)(0.33,-0.33){3}{\tiny.}
\multiput(13.2,56.8)(0.33,-0.32){3}{\tiny.}
\multiput(14.2,55.9)(0.33,-0.31){3}{\tiny.}
\multiput(15.2,54.9)(0.33,-0.31){3}{\tiny.}
\multiput(16.2,54.0)(0.33,-0.30){3}{\tiny.}
\multiput(17.2,53.1)(0.33,-0.29){3}{\tiny.}
\multiput(18.2,52.2)(0.33,-0.29){3}{\tiny.}
\multiput(19.2,51.4)(0.33,-0.28){3}{\tiny.}
\multiput(20.2,50.5)(0.33,-0.28){3}{\tiny.}
\multiput(21.2,49.7)(0.33,-0.27){3}{\tiny.}
\multiput(22.2,48.9)(0.33,-0.27){3}{\tiny.}
\multiput(23.2,48.1)(0.33,-0.26){3}{\tiny.}
\multiput(24.2,47.3)(0.33,-0.26){3}{\tiny.}
\multiput(25.2,46.5)(0.33,-0.25){3}{\tiny.}
\multiput(26.2,45.7)(0.33,-0.25){3}{\tiny.}
\multiput(27.2,45.0)(0.33,-0.24){3}{\tiny.}
\multiput(28.2,44.3)(0.33,-0.24){3}{\tiny.}
\multiput(29.2,43.5)(0.33,-0.23){3}{\tiny.}
\multiput(30.2,42.8)(0.33,-0.23){3}{\tiny.}
\multiput(31.2,42.1)(0.33,-0.23){3}{\tiny.}
\multiput(32.2,41.5)(0.33,-0.22){3}{\tiny.}
\multiput(33.2,40.8)(0.33,-0.22){3}{\tiny.}
\multiput(34.2,40.1)(0.33,-0.21){3}{\tiny.}
\multiput(35.2,39.5)(0.33,-0.21){3}{\tiny.}
\multiput(36.2,38.9)(0.33,-0.21){3}{\tiny.}
\multiput(37.2,38.3)(0.33,-0.20){3}{\tiny.}
\multiput(38.2,37.6)(0.33,-0.20){3}{\tiny.}
\multiput(39.2,37.0)(0.33,-0.20){3}{\tiny.}
\multiput(40.2,36.5)(0.33,-0.19){3}{\tiny.}
\multiput(41.2,35.9)(0.33,-0.19){3}{\tiny.}
\multiput(42.2,35.3)(0.33,-0.19){3}{\tiny.}
\multiput(43.2,34.8)(0.33,-0.18){3}{\tiny.}
\multiput(44.2,34.2)(0.33,-0.18){3}{\tiny.}
\multiput(45.2,33.7)(0.33,-0.18){3}{\tiny.}
\multiput(46.2,33.1)(0.33,-0.17){3}{\tiny.}
\multiput(47.2,32.6)(0.33,-0.17){3}{\tiny.}
\multiput(48.2,32.1)(0.33,-0.17){3}{\tiny.}
\multiput(49.2,31.6)(0.33,-0.16){3}{\tiny.}
\multiput(50.2,31.1)(0.33,-0.16){3}{\tiny.}
\multiput(51.2,30.6)(0.33,-0.16){3}{\tiny.}
\multiput(52.2,30.2)(0.33,-0.16){3}{\tiny.}
\multiput(53.2,29.7)(0.33,-0.15){3}{\tiny.}
\multiput(54.2,29.2)(0.33,-0.15){3}{\tiny.}
\multiput(55.2,28.8)(0.33,-0.15){3}{\tiny.}
\multiput(56.2,28.3)(0.33,-0.15){3}{\tiny.}
\multiput(57.2,27.9)(0.33,-0.14){3}{\tiny.}
\multiput(58.2,27.5)(0.33,-0.14){3}{\tiny.}
\multiput(59.2,27.0)(0.33,-0.14){3}{\tiny.}
\multiput(60.2,26.6)(0.33,-0.14){3}{\tiny.}
\multiput(61.2,26.2)(0.33,-0.13){3}{\tiny.}
\multiput(62.2,25.8)(0.33,-0.13){3}{\tiny.}
\multiput(63.2,25.4)(0.33,-0.13){3}{\tiny.}
\multiput(64.2,25.0)(0.33,-0.13){3}{\tiny.}
\multiput(65.2,24.6)(0.33,-0.13){3}{\tiny.}
\multiput(66.2,24.3)(0.33,-0.12){3}{\tiny.}
\multiput(67.2,23.9)(0.33,-0.12){3}{\tiny.}
\multiput(68.2,23.5)(0.33,-0.12){3}{\tiny.}
\multiput(69.2,23.2)(0.33,-0.12){3}{\tiny.}
\multiput(70.2,22.8)(0.33,-0.12){3}{\tiny.}
\multiput(71.2,22.5)(0.33,-0.11){3}{\tiny.}
\multiput(72.2,22.1)(0.33,-0.11){3}{\tiny.}
\multiput(73.2,21.8)(0.33,-0.11){3}{\tiny.}
\multiput(74.2,21.4)(0.33,-0.11){3}{\tiny.}
\multiput(75.2,21.1)(0.33,-0.11){3}{\tiny.}
\multiput(76.2,20.8)(0.33,-0.11){3}{\tiny.}
\multiput(77.2,20.5)(0.33,-0.10){3}{\tiny.}
\multiput(78.2,20.2)(0.33,-0.10){3}{\tiny.}
\multiput(79.2,19.9)(0.33,-0.10){3}{\tiny.}
\multiput(80.2,19.6)(0.33,-0.10){3}{\tiny.}
\multiput(81.2,19.3)(0.33,-0.10){3}{\tiny.}
\multiput(82.2,19.0)(0.33,-0.10){3}{\tiny.}
\multiput(83.2,18.7)(0.33,-0.09){3}{\tiny.}
\multiput(84.2,18.4)(0.33,-0.09){3}{\tiny.}
\multiput(85.2,18.1)(0.33,-0.09){3}{\tiny.}
\multiput(86.2,17.9)(0.33,-0.09){3}{\tiny.}
\multiput(87.2,17.6)(0.33,-0.09){3}{\tiny.}
\multiput(88.2,17.3)(0.33,-0.09){3}{\tiny.}
\multiput(89.2,17.1)(0.33,-0.09){3}{\tiny.}
\multiput(90.2,16.8)(0.33,-0.08){3}{\tiny.}
\multiput(91.2,16.5)(0.33,-0.08){3}{\tiny.}
\multiput(92.2,16.3)(0.33,-0.08){3}{\tiny.}
\multiput(93.2,16.0)(0.33,-0.08){3}{\tiny.}
\multiput(94.2,15.8)(0.33,-0.08){3}{\tiny.}
\multiput(95.2,15.6)(0.33,-0.08){3}{\tiny.}
\multiput(96.2,15.3)(0.33,-0.08){3}{\tiny.}
\multiput(97.2,15.1)(0.33,-0.08){3}{\tiny.}
\multiput(98.2,14.9)(0.33,-0.08){3}{\tiny.}
\multiput(99.2,14.6)(0.33,-0.07){3}{\tiny.}
\multiput(100.2,14.4)(0.33,-0.07){3}{\tiny.}
\multiput(101.2,14.2)(0.33,-0.07){3}{\tiny.}
\multiput(102.2,14.0)(0.33,-0.07){3}{\tiny.}
\multiput(103.2,13.8)(0.33,-0.07){3}{\tiny.}
\multiput(104.2,13.6)(0.33,-0.07){3}{\tiny.}
\multiput(105.2,13.3)(0.33,-0.07){3}{\tiny.}
\multiput(106.2,13.1)(0.33,-0.07){3}{\tiny.}
\multiput(107.2,12.9)(0.33,-0.07){3}{\tiny.}
\multiput(108.2,12.7)(0.33,-0.06){3}{\tiny.}
\multiput(109.2,12.6)(0.33,-0.06){3}{\tiny.}
\multiput(110.2,12.4)(0.33,-0.06){3}{\tiny.}
\multiput(111.2,12.2)(0.33,-0.06){3}{\tiny.}
\multiput(112.2,12.0)(0.33,-0.06){3}{\tiny.}
\multiput(113.2,11.8)(0.33,-0.06){3}{\tiny.}
\multiput(114.2,11.6)(0.33,-0.06){3}{\tiny.}
\multiput(115.2,11.4)(0.33,-0.06){3}{\tiny.}
\multiput(116.2,11.3)(0.33,-0.06){3}{\tiny.}
\multiput(117.2,11.1)(0.33,-0.06){3}{\tiny.}
\multiput(118.2,10.9)(0.33,-0.06){3}{\tiny.}
\multiput(119.2,10.8)(0.33,-0.06){3}{\tiny.}
\multiput(120.2,10.6)(0.33,-0.05){3}{\tiny.}
\multiput(121.2,10.4)(0.33,-0.05){3}{\tiny.}
\multiput(122.2,10.3)(0.33,-0.05){3}{\tiny.}
\multiput(123.2,10.1)(0.33,-0.05){3}{\tiny.}
\multiput(124.2,9.9)(0.33,-0.05){3}{\tiny.}
\multiput(125.2,9.8)(0.33,-0.05){3}{\tiny.}
\multiput(126.2,9.6)(0.33,-0.05){3}{\tiny.}
\multiput(127.2,9.5)(0.33,-0.05){3}{\tiny.}
\multiput(128.2,9.3)(0.33,-0.05){3}{\tiny.}
\multiput(129.2,9.2)(0.33,-0.05){3}{\tiny.}
\multiput(130.2,9.0)(0.33,-0.05){3}{\tiny.}
\multiput(131.2,8.9)(0.33,-0.05){3}{\tiny.}
\multiput(132.2,8.8)(0.33,-0.05){3}{\tiny.}
\multiput(133.2,8.6)(0.33,-0.05){3}{\tiny.}
\multiput(134.2,8.5)(0.33,-0.04){3}{\tiny.}
\multiput(135.2,8.4)(0.33,-0.04){3}{\tiny.}
\multiput(136.2,8.2)(0.33,-0.04){3}{\tiny.}
\multiput(137.2,8.1)(0.33,-0.04){3}{\tiny.}
\multiput(138.2,8.0)(0.33,-0.04){3}{\tiny.}
\multiput(139.2,7.8)(0.33,-0.04){3}{\tiny.}
\multiput(140.2,7.7)(0.33,-0.04){3}{\tiny.}
\multiput(141.2,7.6)(0.33,-0.04){3}{\tiny.}
\multiput(142.2,7.5)(0.33,-0.04){3}{\tiny.}
\multiput(143.2,7.3)(0.33,-0.04){3}{\tiny.}
\multiput(144.2,7.2)(0.33,-0.04){3}{\tiny.}
\multiput(145.2,7.1)(0.33,-0.04){3}{\tiny.}
\multiput(146.2,7.0)(0.33,-0.04){3}{\tiny.}
\multiput(147.2,6.9)(0.33,-0.04){3}{\tiny.}
\multiput(148.2,6.8)(0.33,-0.04){3}{\tiny.}
\multiput(149.2,6.7)(0.33,-0.04){3}{\tiny.}
\multiput(-1.5,74.3)(0.33,-0.72){3}{\small.}
\multiput(-0.5,72.1)(0.33,-0.70){3}{\small.}
\multiput(0.5,70.1)(0.33,-0.67){3}{\small.}
\multiput(1.5,68.0)(0.33,-0.65){3}{\small.}
\multiput(2.5,66.1)(0.33,-0.62){3}{\small.}
\multiput(3.5,64.2)(0.33,-0.60){3}{\small.}
\multiput(4.5,62.4)(0.33,-0.58){3}{\small.}
\multiput(5.5,60.7)(0.33,-0.56){3}{\small.}
\multiput(6.5,59.0)(0.33,-0.54){3}{\small.}
\multiput(7.5,57.4)(0.33,-0.52){3}{\small.}
\multiput(8.5,55.9)(0.33,-0.50){3}{\small.}
\multiput(9.5,54.4)(0.33,-0.48){3}{\small.}
\multiput(10.5,52.9)(0.33,-0.47){3}{\small.}
\multiput(11.5,51.5)(0.33,-0.45){3}{\small.}
\multiput(12.5,50.2)(0.33,-0.43){3}{\small.}
\multiput(13.5,48.9)(0.33,-0.42){3}{\small.}
\multiput(14.5,47.6)(0.33,-0.40){3}{\small.}
\multiput(15.5,46.4)(0.33,-0.39){3}{\small.}
\multiput(16.5,45.2)(0.33,-0.38){3}{\small.}
\multiput(17.5,44.1)(0.50,-0.55){2}{\small.}
\multiput(18.5,43.0)(0.50,-0.53){2}{\small.}
\multiput(19.5,41.9)(0.50,-0.51){2}{\small.}
\multiput(20.5,40.9)(0.50,-0.49){2}{\small.}
\multiput(21.5,39.9)(0.50,-0.48){2}{\small.}
\multiput(22.5,39.0)(0.50,-0.46){2}{\small.}
\multiput(23.5,38.1)(0.50,-0.44){2}{\small.}
\multiput(24.5,37.2)(0.50,-0.43){2}{\small.}
\multiput(25.5,36.3)(0.50,-0.42){2}{\small.}
\multiput(26.5,35.5)(0.50,-0.40){2}{\small.}
\multiput(27.5,34.7)(0.50,-0.39){2}{\small.}
\multiput(28.5,33.9)(0.50,-0.38){2}{\small.}
\multiput(29.5,33.2)(0.50,-0.36){2}{\small.}
\multiput(30.5,32.4)(0.50,-0.35){2}{\small.}
\multiput(31.5,31.7)(0.50,-0.34){2}{\small.}
\multiput(32.5,31.0)(0.50,-0.33){2}{\small.}
\multiput(33.5,30.4)(0.50,-0.32){2}{\small.}
\multiput(34.5,29.8)(0.50,-0.31){2}{\small.}
\multiput(35.5,29.1)(0.50,-0.30){2}{\small.}
\multiput(36.5,28.5)(0.50,-0.29){2}{\small.}
\multiput(37.5,28.0)(0.50,-0.28){2}{\small.}
\multiput(38.5,27.4)(0.50,-0.27){2}{\small.}
\multiput(39.5,26.9)(0.50,-0.26){2}{\small.}
\multiput(40.5,26.3)(0.50,-0.26){2}{\small.}
\multiput(41.5,25.8)(0.50,-0.25){2}{\small.}
\multiput(42.5,25.3)(0.50,-0.24){2}{\small.}
\multiput(43.5,24.8)(0.50,-0.23){2}{\small.}
\multiput(44.5,24.4)(0.50,-0.23){2}{\small.}
\multiput(45.5,23.9)(0.50,-0.22){2}{\small.}
\multiput(46.5,23.5)(0.50,-0.21){2}{\small.}
\multiput(47.5,23.1)(0.50,-0.21){2}{\small.}
\multiput(48.5,22.7)(0.50,-0.20){2}{\small.}
\multiput(49.5,22.3)(0.50,-0.19){2}{\small.}
\multiput(50.5,21.9)(0.50,-0.19){2}{\small.}
\multiput(51.5,21.5)(0.50,-0.18){2}{\small.}
\multiput(52.5,21.1)(0.50,-0.18){2}{\small.}
\multiput(53.5,20.8)(0.50,-0.17){2}{\small.}
\multiput(54.5,20.4)(0.50,-0.17){2}{\small.}
\multiput(55.5,20.1)(0.50,-0.16){2}{\small.}
\multiput(56.5,19.8)(0.50,-0.16){2}{\small.}
\multiput(57.5,19.4)(0.50,-0.15){2}{\small.}
\multiput(58.5,19.1)(0.50,-0.15){2}{\small.}
\multiput(59.5,18.8)(0.50,-0.15){2}{\small.}
\multiput(60.5,18.5)(0.50,-0.14){2}{\small.}
\multiput(61.5,18.3)(0.50,-0.14){2}{\small.}
\multiput(62.5,18.0)(0.50,-0.13){2}{\small.}
\multiput(63.5,17.7)(0.50,-0.13){2}{\small.}
\multiput(64.5,17.5)(0.50,-0.13){2}{\small.}
\multiput(65.5,17.2)(0.50,-0.12){2}{\small.}
\multiput(66.5,17.0)(0.50,-0.12){2}{\small.}
\multiput(67.5,16.7)(0.50,-0.12){2}{\small.}
\multiput(68.5,16.5)(0.50,-0.11){2}{\small.}
\multiput(69.5,16.3)(0.50,-0.11){2}{\small.}
\multiput(70.5,16.0)(0.50,-0.11){2}{\small.}
\multiput(71.5,15.8)(0.50,-0.11){2}{\small.}
\multiput(72.5,15.6)(0.50,-0.10){2}{\small.}
\multiput(73.5,15.4)(0.50,-0.10){2}{\small.}
\multiput(74.5,15.2)(0.50,-0.10){2}{\small.}
\multiput(75.5,15.0)(0.50,-0.10){2}{\small.}
\multiput(76.5,14.8)(0.50,-0.09){2}{\small.}
\multiput(77.5,14.6)(0.50,-0.09){2}{\small.}
\multiput(78.5,14.4)(0.50,-0.09){2}{\small.}
\multiput(79.5,14.3)(0.50,-0.09){2}{\small.}
\multiput(80.5,14.1)(0.50,-0.08){2}{\small.}
\multiput(81.5,13.9)(0.50,-0.08){2}{\small.}
\multiput(82.5,13.8)(0.50,-0.08){2}{\small.}
\multiput(83.5,13.6)(0.50,-0.08){2}{\small.}
\multiput(84.5,13.4)(0.50,-0.08){2}{\small.}
\multiput(85.5,13.3)(0.50,-0.08){2}{\small.}
\multiput(86.5,13.1)(0.50,-0.07){2}{\small.}
\multiput(87.5,13.0)(0.50,-0.07){2}{\small.}
\multiput(88.5,12.8)(0.50,-0.07){2}{\small.}
\multiput(89.5,12.7)(0.50,-0.07){2}{\small.}
\multiput(90.5,12.6)(0.50,-0.07){2}{\small.}
\multiput(91.5,12.4)(0.50,-0.07){2}{\small.}
\multiput(92.5,12.3)(0.50,-0.06){2}{\small.}
\multiput(93.5,12.2)(0.50,-0.06){2}{\small.}
\multiput(94.5,12.0)(0.50,-0.06){2}{\small.}
\multiput(95.5,11.9)(0.50,-0.06){2}{\small.}
\multiput(96.5,11.8)(0.50,-0.06){2}{\small.}
\multiput(97.5,11.7)(0.50,-0.06){2}{\small.}
\multiput(98.5,11.6)(0.50,-0.06){2}{\small.}
\multiput(99.5,11.4)(0.50,-0.06){2}{\small.}
\multiput(100.5,11.3)(0.50,-0.05){2}{\small.}
\multiput(101.5,11.2)(0.50,-0.05){2}{\small.}
\multiput(102.5,11.1)(0.50,-0.05){2}{\small.}
\multiput(103.5,11.0)(0.50,-0.05){2}{\small.}
\multiput(104.5,10.9)(0.50,-0.05){2}{\small.}
\multiput(105.5,10.8)(0.50,-0.05){2}{\small.}
\multiput(106.5,10.7)(0.50,-0.05){2}{\small.}
\multiput(107.5,10.6)(0.50,-0.05){2}{\small.}
\multiput(108.5,10.5)(0.50,-0.05){2}{\small.}
\multiput(109.5,10.4)(0.50,-0.05){2}{\small.}
\multiput(110.5,10.3)(0.50,-0.04){2}{\small.}
\multiput(111.5,10.3)(0.50,-0.04){2}{\small.}
\multiput(112.5,10.2)(0.50,-0.04){2}{\small.}
\multiput(113.5,10.1)(0.50,-0.04){2}{\small.}
\multiput(114.5,10.0)(0.50,-0.04){2}{\small.}
\multiput(115.5,9.9)(0.50,-0.04){2}{\small.}
\multiput(116.5,9.8)(0.50,-0.04){2}{\small.}
\multiput(117.5,9.8)(0.50,-0.04){2}{\small.}
\multiput(118.5,9.7)(0.50,-0.04){2}{\small.}
\multiput(119.5,9.6)(0.50,-0.04){2}{\small.}
\multiput(120.5,9.5)(0.50,-0.04){2}{\small.}
\multiput(121.5,9.5)(0.50,-0.04){2}{\small.}
\multiput(122.5,9.4)(0.50,-0.03){2}{\small.}
\multiput(123.5,9.3)(0.50,-0.03){2}{\small.}
\multiput(124.5,9.2)(0.50,-0.03){2}{\small.}
\multiput(125.5,9.2)(0.50,-0.03){2}{\small.}
\multiput(126.5,9.1)(0.50,-0.03){2}{\small.}
\multiput(127.5,9.1)(0.50,-0.03){2}{\small.}
\multiput(128.5,9.0)(0.50,-0.03){2}{\small.}
\multiput(129.5,8.9)(0.50,-0.03){2}{\small.}
\multiput(130.5,8.9)(0.50,-0.03){2}{\small.}
\multiput(131.5,8.8)(0.50,-0.03){2}{\small.}
\multiput(132.5,8.7)(0.50,-0.03){2}{\small.}
\multiput(133.5,8.7)(0.50,-0.03){2}{\small.}
\multiput(134.5,8.6)(0.50,-0.03){2}{\small.}
\multiput(135.5,8.6)(0.50,-0.03){2}{\small.}
\multiput(136.5,8.5)(0.50,-0.03){2}{\small.}
\multiput(137.5,8.5)(0.50,-0.03){2}{\small.}
\multiput(138.5,8.4)(0.50,-0.03){2}{\small.}
\multiput(139.5,8.4)(0.50,-0.03){2}{\small.}
\multiput(140.5,8.3)(0.50,-0.02){2}{\small.}
\multiput(141.5,8.3)(0.50,-0.02){2}{\small.}
\multiput(142.5,8.2)(0.50,-0.02){2}{\small.}
\multiput(143.5,8.2)(0.50,-0.02){2}{\small.}
\multiput(144.5,8.1)(0.50,-0.02){2}{\small.}
\multiput(145.5,8.1)(0.50,-0.02){2}{\small.}
\multiput(146.5,8.0)(0.50,-0.02){2}{\small.}
\multiput(147.5,8.0)(0.50,-0.02){2}{\small.}
\multiput(148.5,7.9)(0.50,-0.02){2}{\small.}
\multiput(-1.1,-0.4)(0.60,0.01){5}{\scriptsize.}
\multiput(4.9,-0.3)(0.60,0.03){5}{\scriptsize.}
\multiput(10.9,0.1)(0.60,0.04){5}{\scriptsize.}
\multiput(16.9,0.5)(0.60,0.05){5}{\scriptsize.}
\multiput(22.9,1.0)(0.60,0.05){5}{\scriptsize.}
\multiput(28.9,1.4)(0.60,0.04){5}{\scriptsize.}
\multiput(34.9,1.9)(0.60,0.04){5}{\scriptsize.}
\multiput(40.9,2.3)(0.60,0.04){5}{\scriptsize.}
\multiput(46.9,2.6)(0.60,0.03){5}{\scriptsize.}
\multiput(52.9,2.9)(0.60,0.02){5}{\scriptsize.}
\multiput(58.9,3.1)(0.60,0.02){5}{\scriptsize.}
\multiput(64.9,3.3)(0.60,0.01){5}{\scriptsize.}
\multiput(70.9,3.4)(0.60,0.01){5}{\scriptsize.}
\multiput(76.9,3.5)(0.60,0.00){5}{\scriptsize.}
\multiput(82.9,3.5)(0.60,0.00){5}{\scriptsize.}
\multiput(88.9,3.5)(0.60,-0.00){5}{\scriptsize.}
\multiput(94.9,3.5)(0.60,-0.01){5}{\scriptsize.}
\multiput(100.9,3.4)(0.60,-0.01){5}{\scriptsize.}
\multiput(106.9,3.3)(0.60,-0.01){5}{\scriptsize.}
\multiput(112.9,3.2)(0.60,-0.01){5}{\scriptsize.}
\multiput(118.9,3.1)(0.60,-0.01){5}{\scriptsize.}
\multiput(124.9,3.0)(0.60,-0.01){5}{\scriptsize.}
\multiput(130.9,2.9)(0.60,-0.01){5}{\scriptsize.}
\multiput(136.9,2.7)(0.60,-0.02){5}{\scriptsize.}
\multiput(142.9,2.6)(0.60,-0.02){5}{\scriptsize.}
\multiput(-1.1,72.4)(0.20,-0.49){5}{\scriptsize.}
\multiput(0.9,67.6)(0.25,-0.57){4}{\scriptsize.}
\multiput(2.9,63.1)(0.25,-0.53){4}{\scriptsize.}
\multiput(4.9,58.9)(0.25,-0.50){4}{\scriptsize.}
\multiput(6.9,54.9)(0.25,-0.47){4}{\scriptsize.}
\multiput(8.9,51.2)(0.25,-0.44){4}{\scriptsize.}
\multiput(10.9,47.8)(0.25,-0.41){4}{\scriptsize.}
\multiput(12.9,44.5)(0.25,-0.39){4}{\scriptsize.}
\multiput(14.9,41.5)(0.33,-0.48){3}{\scriptsize.}
\multiput(16.9,38.6)(0.33,-0.46){3}{\scriptsize.}
\multiput(18.9,35.9)(0.33,-0.43){3}{\scriptsize.}
\multiput(20.9,33.4)(0.33,-0.40){3}{\scriptsize.}
\multiput(22.9,31.0)(0.33,-0.38){3}{\scriptsize.}
\multiput(24.9,28.8)(0.33,-0.36){3}{\scriptsize.}
\multiput(26.9,26.7)(0.33,-0.33){3}{\scriptsize.}
\multiput(28.9,24.7)(0.33,-0.31){3}{\scriptsize.}
\multiput(30.9,22.8)(0.33,-0.30){3}{\scriptsize.}
\multiput(32.9,21.1)(0.33,-0.28){3}{\scriptsize.}
\multiput(34.9,19.4)(0.33,-0.26){3}{\scriptsize.}
\multiput(36.9,17.9)(0.33,-0.25){3}{\scriptsize.}
\multiput(38.9,16.4)(0.33,-0.23){3}{\scriptsize.}
\multiput(40.9,15.0)(0.33,-0.22){3}{\scriptsize.}
\multiput(42.9,13.7)(0.33,-0.21){3}{\scriptsize.}
\multiput(44.9,12.5)(0.33,-0.20){3}{\scriptsize.}
\multiput(46.9,11.3)(0.33,-0.19){3}{\scriptsize.}
\multiput(48.9,10.2)(0.33,-0.18){3}{\scriptsize.}
\multiput(50.9,9.2)(0.33,-0.17){3}{\scriptsize.}
\multiput(52.9,8.2)(0.33,-0.16){3}{\scriptsize.}
\multiput(54.9,7.3)(0.33,-0.15){3}{\scriptsize.}
\multiput(56.9,6.4)(0.33,-0.14){3}{\scriptsize.}
\multiput(58.9,5.6)(0.33,-0.13){3}{\scriptsize.}
\multiput(60.9,4.8)(0.50,-0.19){2}{\scriptsize.}
\multiput(62.9,4.1)(0.50,-0.18){2}{\scriptsize.}
\multiput(64.9,3.4)(0.50,-0.17){2}{\scriptsize.}
\multiput(66.9,2.7)(0.50,-0.16){2}{\scriptsize.}
\multiput(68.9,2.1)(0.50,-0.15){2}{\scriptsize.}
\multiput(70.9,1.5)(0.50,-0.14){2}{\scriptsize.}
\multiput(72.9,0.9)(0.50,-0.13){2}{\scriptsize.}
\multiput(74.9,0.4)(0.50,-0.13){2}{\scriptsize.}
\multiput(76.9,-0.1)(0.50,-0.12){2}{\scriptsize.}
\multiput(78.9,-0.6)(0.50,-0.11){2}{\scriptsize.}
\put(66.,29.){$|G|^2-1$}
\put(64.0,8.0){$D$}
\put(17.,21.6){$F$}
\put(19.,4){$E$}
\end{picture}

\vspace{5cm}
FIG. 3.
The noise $D$ (thick curve), linear lower bound $G^2-1$ (thin
curve), and lower bounds $E$ (dotted curve) and $F$ (dashed curve),
versus $x$ for the Cummings-Tavis amplifier at $N=10$, $t= 0.5$.%

\newpage
\vspace{2cm}
\hspace{2cm}
\begin{picture}(100,130)
\put(-.3,-.3){\rule{.5pt}{72pt}}
\put(-.3,-.3){\rule{94pt}{.507pt}}
\multiput(-0.,9.95)(0,10){7}{\rule{90pt}{.1pt}}
\multiput(9.95,0. )(10,0){9}{\rule{.1pt}{70pt}}
\put(-9,75){\small $D_{0.5}(x),E,F $}
\put(-6, 57){3}
\put(-6, 37){2}
\put(-6, 17){1}
\put(-6, -2){0}
\put( -2,-8){0}
\put( 18,-8){1}
\put( 38,-8){2}
\put( 58,-8){3}
\put( 78,-8){4}
\put( 88,-10){$|G_{0.5}(x)|^2$}
\put(20,0){\line(1,1){74}}
\multiput(91.2,74.3)(-0.50,-0.85){5}{.}
\multiput(88.6,70.1)(-0.48,-0.79){5}{.}
\multiput(86.2,66.1)(-0.58,-0.92){4}{.}
\multiput(83.9,62.4)(-0.56,-0.85){4}{.}
\multiput(81.7,59.0)(-0.53,-0.79){4}{.}
\multiput(79.6,55.9)(-0.51,-0.74){4}{.}
\multiput(77.5,52.9)(-0.49,-0.69){4}{.}
\multiput(75.5,50.2)(-0.47,-0.64){4}{.}
\multiput(73.6,47.6)(-0.61,-0.79){3}{.}
\multiput(71.8,45.2)(-0.58,-0.74){3}{.}
\multiput(70.1,43.0)(-0.56,-0.69){3}{.}
\multiput(68.4,40.9)(-0.54,-0.65){3}{.}
\multiput(66.8,39.0)(-0.52,-0.60){3}{.}
\multiput(65.2,37.2)(-0.50,-0.56){3}{.}
\multiput(63.7,35.5)(-0.48,-0.53){3}{.}
\multiput(62.2,33.9)(-0.47,-0.49){3}{.}
\multiput(60.8,32.4)(-0.45,-0.46){3}{.}
\multiput(59.5,31.0)(-0.65,-0.65){2}{.}
\multiput(58.2,29.8)(-0.63,-0.61){2}{.}
\multiput(57.0,28.5)(-0.60,-0.57){2}{.}
\multiput(55.7,27.4)(-0.58,-0.53){2}{.}
\multiput(54.6,26.3)(-0.56,-0.50){2}{.}
\multiput(53.5,25.3)(-0.54,-0.47){2}{.}
\multiput(52.4,24.4)(-0.52,-0.44){2}{.}
\multiput(51.3,23.5)(-0.51,-0.42){2}{.}
\multiput(50.3,22.7)(-0.49,-0.39){2}{.}
\multiput(49.3,21.9)(-0.47,-0.37){2}{.}
\multiput(48.4,21.1)(-0.46,-0.35){2}{.}
\multiput(47.5,20.4)(-0.44,-0.33){2}{.}
\multiput(46.6,19.8)(-0.43,-0.31){2}{.}
\multiput(45.7,19.1)(-0.41,-0.30){2}{.}
\multiput(44.9,18.5)(-0.40,-0.28){2}{.}
\multiput(44.1,18.0)(-0.39,-0.26){2}{.}
\multiput(43.3,17.5)(-0.37,-0.25){2}{.}
\multiput(42.6,17.0)(-0.36,-0.24){2}{.}
\multiput(41.9,16.5)(-0.35,-0.23){2}{.}
\multiput(41.2,16.0)(-0.34,-0.21){2}{.}
\multiput(40.5,15.6)(-0.33,-0.20){2}{.}
\multiput(39.8,15.2)(-0.32,-0.19){2}{.}
\multiput(39.2,14.8)(-0.31,-0.18){2}{.}
\multiput(38.6,14.4)(-0.30,-0.18){2}{.}
\multiput(38.0,14.1)(-0.29,-0.17){2}{.}
\multiput(37.4,13.8)(-0.28,-0.16){2}{.}
\multiput(36.8,13.4)(-0.27,-0.15){2}{.}
\multiput(36.3,13.1)(-0.53,-0.29){1}{.}
\multiput(35.8,12.8)(-0.51,-0.28){1}{.}
\multiput(35.2,12.6)(-0.50,-0.27){1}{.}
\multiput(34.7,12.3)(-0.48,-0.26){1}{.}
\multiput(34.3,12.0)(-0.47,-0.24){1}{.}
\multiput(33.8,11.8)(-0.45,-0.23){1}{.}
\multiput(33.3,11.6)(-0.44,-0.22){1}{.}
\multiput(32.9,11.3)(-0.43,-0.21){1}{.}
\multiput(32.5,11.1)(-0.42,-0.21){1}{.}
\multiput(32.0,10.9)(-0.40,-0.20){1}{.}
\multiput(31.6,10.7)(-0.39,-0.19){1}{.}
\multiput(31.3,10.5)(-0.38,-0.18){1}{.}
\multiput(30.9,10.3)(-0.37,-0.18){1}{.}
\multiput(30.5,10.2)(-0.36,-0.17){1}{.}
\multiput(30.1,10.0)(-0.35,-0.16){1}{.}
\multiput(29.8,9.8)(-0.34,-0.16){1}{.}
\multiput(29.5,9.7)(-0.33,-0.15){1}{.}
\multiput(29.1,9.5)(-0.32,-0.14){1}{.}
\multiput(28.8,9.4)(-0.31,-0.14){1}{.}
\multiput(28.5,9.2)(-0.30,-0.13){1}{.}
\multiput(28.2,9.1)(-0.29,-0.13){1}{.}
\multiput(27.9,9.0)(-0.29,-0.12){1}{.}
\multiput(27.6,8.9)(-0.28,-0.12){1}{.}
\multiput(27.3,8.7)(-0.27,-0.11){1}{.}
\multiput(27.1,8.6)(-0.26,-0.11){1}{.}
\multiput(26.8,8.5)(-0.26,-0.11){1}{.}
\multiput(26.5,8.4)(-0.25,-0.10){1}{.}
\multiput(26.3,8.3)(-0.24,-0.10){1}{.}
\multiput(26.0,8.2)(-0.24,-0.09){1}{.}
\multiput(25.8,8.1)(-0.23,-0.09){1}{.}
\multiput(25.6,8.0)(-0.22,-0.09){1}{.}
\multiput(25.4,7.9)(-0.22,-0.08){1}{.}
\multiput(25.1,7.9)(-0.21,-0.08){1}{.}
\multiput(24.9,7.8)(-0.21,-0.08){1}{.}
\multiput(24.7,7.7)(-0.20,-0.07){1}{.}
\multiput(24.5,7.6)(-0.19,-0.07){1}{.}
\multiput(24.3,7.6)(-0.19,-0.07){1}{.}
\multiput(24.1,7.5)(-0.18,-0.06){1}{.}
\multiput(24.0,7.4)(-0.18,-0.06){1}{.}
\multiput(23.8,7.4)(-0.17,-0.06){1}{.}
\multiput(23.6,7.3)(-0.17,-0.05){1}{.}
\multiput(23.4,7.3)(-0.16,-0.05){1}{.}
\multiput(23.3,7.2)(-0.16,-0.05){1}{.}
\multiput(23.1,7.2)(-0.16,-0.05){1}{.}
\multiput(23.0,7.1)(-0.15,-0.04){1}{.}
\multiput(22.8,7.1)(-0.15,-0.04){1}{.}
\multiput(22.7,7.0)(-0.14,-0.04){1}{.}
\multiput(22.5,7.0)(-0.14,-0.03){1}{.}
\multiput(22.4,7.0)(-0.14,-0.03){1}{.}
\multiput(22.2,6.9)(-0.13,-0.03){1}{.}
\multiput(22.1,6.9)(-0.13,-0.03){1}{.}
\multiput(22.0,6.9)(-0.13,-0.02){1}{.}
\multiput(21.8,6.9)(-0.12,-0.02){1}{.}
\multiput(21.7,6.8)(-0.12,-0.02){1}{.}
\multiput(21.6,6.8)(-0.12,-0.02){1}{.}
\multiput(21.5,6.8)(-0.11,-0.01){1}{.}
\multiput(91.2,72.3)(-0.42,-0.81){3}{.}
\multiput(87.4,65.2)(-0.40,-0.74){3}{.}
\multiput(83.9,58.8)(-0.37,-0.67){3}{.}
\multiput(80.6,52.9)(-0.35,-0.61){3}{.}
\multiput(77.5,47.7)(-0.33,-0.55){3}{.}
\multiput(74.6,42.9)(-0.47,-0.75){2}{.}
\multiput(71.8,38.5)(-0.44,-0.68){2}{.}
\multiput(69.2,34.5)(-0.42,-0.62){2}{.}
\multiput(66.8,30.9)(-0.39,-0.57){2}{.}
\multiput(64.4,27.6)(-0.37,-0.52){2}{.}
\multiput(62.2,24.6)(-0.35,-0.47){2}{.}
\multiput(60.2,21.8)(-0.33,-0.43){2}{.}
\multiput(58.2,19.3)(-0.32,-0.39){2}{.}
\multiput(56.3,17.0)(-0.30,-0.36){2}{.}
\multiput(54.6,14.9)(-0.28,-0.33){2}{.}
\multiput(52.9,13.0)(-0.27,-0.30){2}{.}
\multiput(51.3,11.2)(-0.26,-0.28){2}{.}
\multiput(49.8,9.6)(-0.24,-0.26){2}{.}
\multiput(48.4,8.1)(-0.23,-0.23){2}{.}
\multiput(47.0,6.7)(-0.44,-0.43){1}{.}
\multiput(45.7,5.5)(-0.42,-0.40){1}{.}
\multiput(44.5,4.3)(-0.40,-0.36){1}{.}
\multiput(43.3,3.3)(-0.38,-0.34){1}{.}
\multiput(42.2,2.3)(-0.36,-0.31){1}{.}
\multiput(41.2,1.4)(-0.34,-0.28){1}{.}
\multiput(40.1,0.6)(-0.33,-0.26){1}{.}
\multiput(39.2,-0.2)(-0.31,-0.24){1}{.}
\multiput(38.3,-0.9)(-0.30,-0.22){1}{.}
\multiput(37.4,-1.6)(-0.28,-0.20){1}{.}
\multiput(91.2,-0.5)(-1.05,0.02){8}{.}
\multiput(79.6,-0.2)(-1.14,0.08){6}{.}
\multiput(70.1,0.5)(-0.94,0.09){6}{.}
\multiput(62.2,1.3)(-0.93,0.10){5}{.}
\multiput(55.7,2.0)(-0.98,0.10){4}{.}
\multiput(50.3,2.6)(-0.82,0.08){4}{.}
\multiput(45.7,3.0)(-0.93,0.07){3}{.}
\multiput(41.9,3.3)(-0.79,0.03){3}{.}
\multiput(38.6,3.4)(-0.67,0.01){3}{.}
\multiput(35.8,3.4)(-0.86,-0.02){2}{.}
\multiput(33.3,3.4)(-0.74,-0.05){2}{.}
\multiput(31.3,3.2)(-0.64,-0.06){2}{.}
\multiput(29.5,3.0)(-0.56,-0.08){2}{.}
\multiput(27.9,2.8)(-0.48,-0.09){2}{.}
\multiput(26.5,2.6)(-0.42,-0.09){2}{.}
\multiput(25.4,2.3)(-0.37,-0.09){2}{.}
\multiput(24.3,2.0)(-0.32,-0.09){2}{.}
\multiput(23.4,1.8)(-0.56,-0.18){1}{.}
\multiput(22.7,1.5)(-0.49,-0.17){1}{.}
\put(61, 4){$E$}
\put(68,25){$F$}
\put(15, 8){$D$}
\end{picture}

\vspace{5cm}
FIG. 4.
The noise $D$ and all bounds versus $|G|^2$ under the same conditions.%

\end{document}